\documentclass{aa}

\usepackage{graphicx}
\usepackage{txfonts}
\usepackage{color}
\newcommand{\msol}{{\rm M}_{\odot}}
\newcommand{\rsol}{{\rm R}_{\odot}}

\newcommand{\onvire}[1]{}
\newcommand{\eq}[1]{Eq.~(\ref{#1})}
\newcommand{\beq}{\begin{equation}}
\newcommand{\eeq}{\end{equation}}

\usepackage{color}
\usepackage{epsfig}
\usepackage{amssymb}
\usepackage{graphicx,natbib}
\begin{document}
   \title{The radial structure of protostellar accretion disks: influence of jets} 

   \author{C. Combet
          \inst{1}
          \and
          J. Ferreira\inst{1}
          }

   \offprints{C. Combet}

   \institute{Laboratoire d'Astrophysique de Grenoble (LAOG), UJF/CNRS,
              BP 53, 38041 Grenoble Cedex 9, France\\
              \email{ccombet@obs.ujf-grenoble.fr}
         }

   \date{Received \today; accepted }

 
  \abstract
   {The radial structure of accretion disks is a fundamental issue regarding star and 
planet formation. Many theoretical studies, focussing on different aspects such as e.g.
disk emissivity or ionization, have been conducted in the context of the Standard 
Accretion Disk (SAD) model, where no jet is present.
}
   {We wish to calculate the structure of YSO accretion disks in an 
approach that takes into account the presence of the protostellar jets. The 
radial structure of these Jet Emitting Disks (JED) should then be compared 
to that of standard accretion disks.}
   {The analytical treatment used in this work is very similar to that of
standard accretion disks but is using the parameter space of Magnetised 
Accretion-Ejection Structures that include the jet torque on the underlying disk.
In this framework, the analytical expressions of key quantities, such as mid-plane temperatures,
surface densities or disk aspect ratio are derived.}
   {It is found that JEDs present a structure very different from the SADs and 
that can be observationally tested. The implications on planet formation in the inner 
regions of accretion disks are briefly discussed.  We also supply sets of analytical 
formulae, valid in different opacity regimes, for the disk quantities. These expressions 
can be readily used for any work where the disk structure is needed as an input for the model.}
   {}

\authorrunning{Combet \& Ferreira}
\titlerunning{Jets and accretion disk structure}

   \maketitle
%

\section{Introduction}
 
Accretion disks are ubiquitous in the Universe. In particular, 
they are found in Active Galactic Nuclei (AGN),
around stellar black holes (X-ray binaries) and in Young Stellar Objects (YSO).
From the theoretical point of view, accretion disks have been extensively studied 
in the context of the Standard Accretion Disk model (hereafter SAD,  
\citealt{1972A&A....21....1P,1973A&A....24..337S,1981ARA&A..19..137P}), be it for AGN or YSO.
These early studies were conducted before the discovery of jets and focused on 
the physics of accretion only. Many refinements have 
been included to the initial SAD approach over the last thirty years, 
but the basic idea is still that of the first seminal papers: the gas inward motion is ensured 
by the radial turbulent transport 
of angular momentum from the inner to the outer parts of the disk. To date, the magneto-rotational
instability (MRI, \citealp{1991ApJ...376..214B}) is the best candidate to provide and 
sustain the level of turbulence required in accretion disks.
For the specific case of YSO---which is our main concern in this work---the standard
theory of accretion disk has been widely used to determine the radial and vertical structure 
of accretion disks. This is a fundamental issue if one is to understand how star form but also the 
initial conditions of planet formation and migration. 

However, after a few decades of observations, it has appeared very clearly 
that disk accretion
onto a central object and bipolar ejections cannot be disentangled. 
Very briefly, accretion is believed to power the jets which, in turn, 
vertically remove part of the disk 
angular momentum allowing accretion to proceed. This
accretion-ejection picture is observed on many astrophysical scales: jets
are present in AGN, microquasars, YSO and have more recently been observed
emerging from brown dwarfs \citep{1998ARA&A..36..539F,1999ARA&A..37..409M,2007prpl.conf..215B,
2005Natur.435..652W}. Despite the advances in the standard theory of 
accretion disk, the latter does not provide any explanation to the production of jets.

In this work, we will focus on the specific case of YSO. 
About 30\% of T~Tauri stars (Class 2 objects) present bipolar ejection. 
This percentage increases to 100\% for Class 0 objects, the earliest stage of 
star formation. Several models have 
been developed to explain the jets seen in T~Tauri stars. Stellar winds have been invoked
\citep{2002A&A...389.1068S} and may be present in the inner parts of the jets: 
however, such winds cannot sustain
the observed mass loss rates and cannot therefore be the main engine of the jets 
\citep{2006A&A...453..785F}.
To date, two accretion powered wind models exist:  i) the X-wind model
\citep[e.g.][]{2000prpl.conf..789S}
and the ii) extended disk wind model 
(e.g. \citealp{1986ApJ...301..571P,1993ApJ...410..218W, 1993A&A...276..625F}). Both models
are based on the same mechanism, the so-called magneto-rotational launching (originally
developed by \citealp{1982MNRAS.199..883B} for the case of AGN) and only differ by
the origin and configuration of the magnetic field threading the disk and 
the size of the launching region. It will require higher
angular resolution observations to have a definite answer regarding the process(es) at play.
However, \citet{2006A&A...453..785F} have gathered indirect evidence (from
jet rotation velocities) that appears to favor extended disk wind theories.

 If the jets are indeed accretion powered---which seems to be the case
from our present knowledge---then the jets must affect the structure of the underlying
region of the disk that is powering them. Hence, the SAD model cannot be used in this
region where most of the angular momentum is transported away vertically, 
in the jets. In this work, we calculate the radial structure of a Jet Emitting Disk (JED) and compare it to that of a SAD. The paper is organized as follows:
\begin{itemize}
\item In Sec.~{\ref{sec:MAES}}, we briefly
present the framework of the Magnetized Accretion Ejection Structures (MAES). 
This extended disk
wind model is a self-consistent description of the accretion disk and the jet it powers.
This approach allows us to quantify the effect of the jet on the disk. 
\item Using the key parameters of the previous model, the main equations used to 
calculate the JED/SAD structure are provided in Sec.~\ref{sec:basic}.
\item The results, and in particular a comparison between the two 
(jet emitting and standard) types of accretion disks, are presented in Sec.~\ref{sec:results}.
\item Before concluding, Sec.~\ref{sec:discussion} raises a few issues and focusses on
the possible implications of the existence of JED with regard to planet formation and migration. 
\end{itemize}


\section{Jet Emitting Disks: The MAES framework \label{sec:MAES}}
The Magnetised Accretion Ejection Structures model \citep{1993A&A...276..625F} 
has been developed so as to treat consistently
both the accretion disk and the jet it generates. The idea is the same as in
earlier studies of magneto-centrifugally launched disk winds \citep{1982MNRAS.199..883B}. However,
in the MAES, the solution starts from the midplane of the resistive MHD disk and evolves outwards
in the ideal MHD wind/jet. This differs drastically from other studies where the disk was only 
treated as a boundary condition, hence forbidding any precise quantification of the effect of the MHD 
wind on the disk.

\subsection{Properties relevant to this work\label{subsec:MAESprop}}

It would be lengthy but also irrelevant to present the MAES model in great detail in this 
paper. Many papers have dealt with the subject, from both analytical and numerical point of views, 
and we refer the reader to these papers for further details (e.g. \citealp{1997A&A...319..340F,
2000A&A...353.1115C,2000A&A...361.1178C,ferreira-2002,2004ApJ...601L.139F,2002ApJ...581..988C,2004ApJ...601...90C,
2007A&A...469..811Z}).
Instead, we give hereafter the few key elements of the model that are important to our work.

In a resistive MAES disk, both angular momentum and magnetic field are transported using an alpha prescription. To that end, a local turbulent resistivity ($\nu_m$ magnetic diffusivity) is supplemented to the \emph{usual}  turbulent viscosity  $\nu_v$ (used in the standard theory, \citealt{1973A&A....24..337S}):
\begin{eqnarray}
\nu_v&=&\alpha_v\Omega_K h^2\;,\\\nonumber
\nu_m&=&\alpha_m v_A h\;,
\end{eqnarray}
with $\Omega_K$ the Keplerian rotation frequency, $v_A$ the Alfv\`en velocity and $h$ the disk
half thickness. 

Accretion solely depends on the removal of the disk angular momentum.
A relevant quantity is then the ratio $\Lambda$ of the jet torque to the viscous torque. This ratio
characterizes the dominant agent responsible for the extraction of angular momentum. For a SAD,
no jet is present: angular momentum is only radially transported away by the 
turbulent viscosity and $\Lambda=0$. However, in a JED, angular momentum is vertically
transported in the wind, along the magnetic field lines, and $\Lambda>0$. It has been shown in previous
work that steady ejection requires $\Lambda\sim 1/\epsilon$, with $\epsilon=h/r$ the disk aspect ratio 
\citep{ferreira-2002}. In 
general MAES solutions present large values of $\Lambda$: this emphasizes the fact that, from the moment a jet is launched, most of the energy and angular momentum will be evacuated in the jets (see Appendix~\ref{app:bilan}).

Concerning the transport of matter, the accretion rate in MAES solutions follows 
$\dot M_a\propto r^\xi$, where
$\xi$ is the ejection efficiency. One of the results of the MAES is that the latter is found to lie
in the range [$10^{-3},5\times 10^{-1}$] in order to steadily provide super-Alfv\'enic jets. Mass conservation in a MAES ranging from the inner radius $r_{in}$ to the outer radius $r_{J}$ reads $\dot M(r_J)=2\dot M_{\rm jet}+\dot M(r_{in})$ from which
one gets,
\beq
\frac{2\dot M_J}{\dot M(r_J)}=1-\left(\frac{r_{in}}{r_J}\right)^\xi \simeq \xi \ln \frac{r_J}{r_{in}}\;.
\eeq
The amount of matter being ejected thus depends on the ejection index, but also
on the extension of the jet emitting disk. Hence, with the typical values $r_{in}=0.04$~AU,
$r_{J}=0.5$~AU and $\xi=0.05$, $\sim 10$\% of the mass will escape in the jets, in agreement with current observational estimates (for more details see \citealt{2006A&A...453..785F}).

The more angular momentum is removed the larger the accretion velocity $u_0$. The sonic Mach number in
the disk $m_s\equiv u_0/\Omega_K h$ can be rewritten under the form 
\beq
m_s\equiv \frac{u_0}{\Omega_K h}=\alpha_v\epsilon+2q\mu=\alpha_v\epsilon(1+\Lambda)\;,
\label{eq:ms}
\eeq
where $\alpha_v\epsilon$ denotes the effect of standard transport and $2q\mu$ 
the specific contribution of the magnetic torque. 
The magnetization\footnote{It is directly linked to the usual plasma beta parameter by $\mu=2/\beta$.}
$\mu=B^2/\mu_0 P$ measures the strength of the magnetic field in the disk and  $q= \mu_o J_r h/B_z$ is the normalized radial electric current density flowing at the disk midplane. This last parameter measures the magnetic shear as it provides an estimate of the toroidal magnetic field component at the disk surface, namely $B_\phi^+ \simeq -q B_z$.

In a SAD, $m_s=\alpha_v\epsilon$, with typically $\alpha_v=10^{-2}$ and $\epsilon=h/r<1$, so that 
the accretion velocity is largely subsonic. However, the situation in a JED is very different as 
MAES solutions present high accretion velocities with $m_s\sim 1$.
This is because steady-state MAES solutions are generally found close to
equipartition with $\mu\in[0.1-1]$ and $q$ of the order of unity
\citep{1995A&A...295..807F,1997A&A...319..340F}. In turn, $q\sim 1$ is only possible for a large level of turbulence, namely $\alpha_m\sim 1$: for smaller values of the magnetic diffusivity, the toroidal field is much larger and the vertical equilibrium is no more possible. As a consequence the torques ratio writes $\Lambda=  2q\mu/\alpha_v \epsilon \simeq 1/\epsilon$, where we assumed for simplicity  $\alpha_v = \alpha_m$. Note that if the viscous parameter $\alpha_v$ is much smaller than unity, $\Lambda$ becomes even larger.

\subsection{JED---SAD transition\label{subsec:transition}}

\begin{figure}
\begin{center}
\includegraphics[clip=, width=9cm]{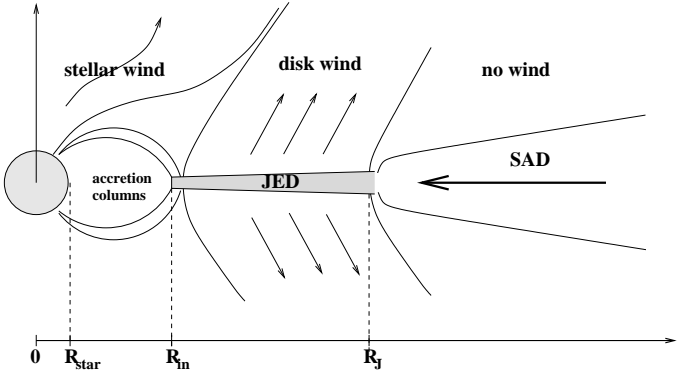}
\caption{Sketch of the accretion configuration suggested in this work. The accretion disk
is constituted by a Standard Accretion Disk in the outer part whereas the inner
part is occupied by a Jet Emitting Disk. See text for details.
For completeness, accretion columns onto the central object and a possible
stellar wind are also represented, although they are not considered
 in the present study. \label{fig:schema}}
\end{center}
\end{figure}

A schematic representation of the accretion structure we consider here is represented in 
Fig.~\ref{fig:schema}. The outer parts of the disk have the characteristics of a Standard Accretion 
Disk whereas the inner part is occupied by a Jet Emitting Disk. The latter has the properties
of the MAES as detailed above. The magnetization $\mu$ provides the criterium for the transition 
between the two types of disk\footnote{Such a picture has been also successfully applied to X-ray binaries 
\citealp{2006A&A...447..813F}.}. Let us examine this point. 

As said before, a SAD assumes the presence of a turbulent angular momentum transport such that the effective Reynolds number ${\cal R}_e = r u_0/\nu_v$ is of order unity. The usual assumption is then a comparable magnetic Reynolds number  ${\cal R}_m = r u_0/\nu_m$, namely a magnetic Prandtl number of order unity (see e.g. \citealp{1996ApJ...473..403H}). 
If one considers the presence of a large scale magnetic field threading the disc, such a small value of  ${\cal R}_m$ translates into straight field lines, i.e. $B_z \gg B_r^+$. Now, the radial distribution $B_z(r)$ is provided by the induction equation which describes the interplay between advection and diffusion. 
Since there is no significant bending in a SAD, this equation writes in steady state
\begin{equation}
\nu_m \frac{\partial B_z}{\partial r} \simeq u_r B_z
\end{equation}
with the obvious  {\it exact} solution 
\begin{equation}
B_z \propto  r^{- {\cal R}_m}
\label{eq:Bz}
\end{equation}
Any large scale magnetic field will thus be naturally increasing towards the center in a SAD, as long as MHD prevails of course. Apart from triggering the MRI, the dynamical importance of this field is in fact measured by the disk magnetization $\mu(r)$. We thus need to evaluate as well the radial distribution of the total pressure $P \simeq \rho \Omega_K h^2$. It can be written as
\begin{equation}
P =  \frac{\dot M_a \Omega_K}{4 \pi r m_s}   \propto  r^{-5/2}
\end{equation} 
where both  the accretion rate $\dot M_a$  and $m_s= \alpha_v \epsilon$ have been assumed constant\footnote{Note that $\epsilon$ is only slowly varying with the radius in a SAD around a protostar so that $h \propto r$ is a good approximation.} Using Eq.~(\ref{eq:Bz}), we obtain 
\begin{equation}
\mu \propto r^{-\delta} \, \, \mbox{      with      }\, \,  \delta= 2 {\cal R}_m - 5/2 
\end{equation}
Strictly speaking, the magnetization will increase towards the center whenever ${\cal R}_m> 5/4$. Thus, unless the magnetic Prandtl number is significantly smaller than unity, this is a condition that is certainly automatically satisfied in SADs. However, making a transition to a JED requires a large value of $\mu$, namely $\mu \sim 1$: a JED will therefore exist only below a transition radius $r_J$ such that $\mu(r_J) \simeq 1$. 
Below this radius, the radial dependency of the vertical magnetic field is different than that given by Eq.(\ref{eq:Bz}). Indeed, a JED requires a field close to equipartition throughout all its extent so that one gets $B_z \propto r^{-5/4+ \xi/2}$ ($\xi$ is the ejection efficiency).

Not all accretion disks might achieve such a transition as it depends on the magnetization at the 
disk outer edge. This outer boundary condition is itself reminiscent of the history of the 
protostellar system, that is of both the magnetization of the parent cloud and the subsequent 
collapse. It is interesting that, indeed, not all young stars have detectable jets. In our picture, 
that would be explained by the lack of a JED in the inner accretion disk. Using a sample of CTTS, 
\citet{2004A&A...425..973M} found that CTTS are oriented randomly with respect to the local 
interstellar field. This may indicate that interstellar magnetic fields play no strong role 
in enforcing the direction of the final (i.e. stellar) angular momentum. However, sources with 
strong outflows do have disks mostly perpendicular to the field (i.e. jets are aligned to it 
as first found by \citet{1986ApJS...62...39S}), whereas sources with no jet detected are parallel. 
That could be a hint that only objects with disks perpendicular to the interstellar magnetic 
field give birth to magnetized central regions, namely JEDs.

For the purpose of this paper, we will treat $r_J$ as a free parameter and compute the radial 
structure of JEDs. It is noteworthy that $r_J$ might be observationally determined by measuring 
the angular velocity in jets 
\citep{2002ApJ...576..222B,2004A&A...416L...9P}. 
Taking kinematic constraints from several T~Tauri jets, \citet{2006A&A...453..785F} found
 that $r_J$ would typically range between 0.2 to a few astronomical units. However, 
these values must be considered with caution as they might represent upper limits only. 
Furthermore,
we restrict ourselves to the optically thick case in this preliminary study. This is justified
as most CTTS show optically thick inner disks. Nevertheless, one has to bear in mind the existence of
transitional disks \citep{2002ApJ...568.1008C,2005ApJ...621..461D,2007A&A...471..173R}: the spectral 
energy distributions of these disks suggest the presence of large optically thin inner holes, 
that are generally explained by invoking planet growth in 
these inner regions. An optically thin JED could be an 
alternative/complementary interpretation to the large inner gaps they display and this hypothesis
deserves further investigation.

There are also some spectroscopic indications that gas is settled in the innermost hot disk 
regions where dust has been sublimated \citep{2007arXiv0704.1841N}. However, we will not 
consider line emission in this work as these inner regions are probably strongly affected 
by the stellar magnetosphere. We will therefore use for the disk inner radius $r_{in}$ an 
arbitrary but representative value of the disk truncation radius, $r_{in}= 0.04$ AU. 

\section{Calculation of the disk structure\label{sec:basic}}

We are interested in deriving the radial structure of an accretion disk, be it a SAD or JED.
The advantage of the approach used in this work is that both types of disks are described 
using the same formalism: the only difference lies in the values of the MAES parameter 
$\Lambda$, thus $m_s$.  We make the assumption of a geometrically thin,
optically thick steady-state disk, rotating at Keplerian velocity, i.e. with a frequency 
$\Omega_K=\sqrt{GM_\star/r^3}$. We also assume that the accretion rate $\dot M_a$ does not
depend on the distance to the central object. This is justified since solutions to 
the MAES problem give $\dot M_a \propto r^\xi$, with $\xi \sim 10^{-3}$-- $5\times 10^{-1}$
\citep{2000A&A...361.1178C}.
We also assume the gas and the dust to be well coupled, at the same temperature, and that the 
mixture behaves as an ideal gas.

\subsection{Basic equations}
We restrict ourselves to the simple case of a steady state accretion disk. The calculation of 
the disk structure relies on the equality of the cooling and heating terms.
The latter can be written in a generic form as
\beq
Q^+=f\times\frac{G M_\star \dot M_a}{8\pi r^3}
\label{eq:q+}
\eeq
where $f$ represents the fraction of gravitational potential energy that is used to heat the gas. 
For this work, the gas is only locally heated by viscous effects and any other source of heating, 
such as irradiation from the star\footnote{This hypothesis is further discussed in Sec.
\ref{subsec:irrad}.}  or dust-gas collisional heating  (occurring when $T_{\rm dust}\neq T_{\rm gas}$) are discarded. In the MAES context, it has been  shown (see Appendix~\ref{app:bilan}) that
\beq
f=\frac{1}{1+\Lambda}\;.
\label{eq:f}
\eeq
\begin{itemize}
\item For a SAD, $\Lambda=0$ and $f_{\rm SAD}=1$: in that case, the viscosity is responsible 
for converting all the mechanical energy into heat.
\item For a steady-state JED, however, $\Lambda\sim\epsilon^{-1}\gg 1$ and $f_{\rm JED}\ll 1$: 
as mentioned earlier, when a jet is present, most of the disk energy leaves in the jet as an MHD Poynting flux. Only a small fraction of that energy remains in the disk to be converted into heat and radiated away. 
\end{itemize}

As for the cooling, we assume that the disk radiates like a black body with an effective temperature
$T_{\rm eff}$, which leads to \citep{1990ApJ...351..632H}
\beq
Q^-=\sigma T_{\rm eff}^4\approx \frac{3}{8\tau} \sigma T_0^4\;,
\label{eq:q-}
\eeq
where the optical depth $\tau\approx \kappa \rho_0 h$ links the effective temperature 
$T_{\rm eff}$ to the mid-plane temperature $T_0$ \emph{via} the opacity of the gas $\kappa$. 
This link between effective and central temperatures holds only if energy transport 
is done vertically by photon diffusion. \citet{1998ApJ...500..411D} showed that radiative 
engery transport was indeed the dominant mechanism in SAD, with respect to convection or
turbulent transport. However, the level of turbulence required in a JED ($\alpha_m \sim 1$) 
is much larger than that usually assumed in a SAD ($\alpha_v \sim 10^{-2}$) so that 
the results of \citet{1998ApJ...500..411D} may not apply here. Nevertheless, 
without any reliable expression for a turbulent energy flux, we restrict ourselves to 
the usual approximation.
For the opacity, we adopt the standard $\kappa=\bar\kappa \rho_0^a T_0^b$ form, where 
$\bar\kappa$, $a$ and $b$ have to be adjusted regarding the dominant 
coolant present in the gas. We use the values given in \citet{1994ApJ...427..987B} that 
includes eight opacity regimes, from dust dominated cooling to electron scattering 
cooling\footnote{In Appendix~\ref{app:formules}, we give the general expressions 
of the radial structure of a JED, keeping $\bar\kappa$, $a$ and $b$ as free parameters.}
(although this regime is not reached by the  temperatures at play in YSO disks). 
 
The Bell \& Lin opacity adjustments can be used to derive the disk structure analytically, 
which is not the case with tabulated results. The value of the opacity of the gas at a 
given temperature and density is a fundamental quantity regarding the calculation of an 
accretion disk structure and should be evaluated carefully. Many authors have numerically 
calculated Rosseland mean opacities, using different models 
\citep{1994ApJ...437..879A,1996A&A...311..291H,2003A&A...410..611S} and the Bell \& Lin 
prescription deviates from more refined approaches around 1500--1800~K, 
where it significantly underestimates $\kappa$ \citep{2003A&A...410..611S}. 
Thus, to be consistent, we stopped our disk calculations whenever the central temperature 
reached this value. For all JED models, the inner disk edge 
$r_{\rm in}$ is achieved at lower temperatures (see Fig.~\ref{fig:tableau_jed}). 
However, SAD models are affected and this is the reason why, in that case, 
the disk inner radius is actually larger than the chosen $r_{in}$ (see also 
\citealp{1999ApJ...521..823P}). 
Hence, more refined SAD calculations should be performed at these inner radii, probably taking into 
account dust sublimation and line emission from the gas component. This is however beyond the 
scope of the present paper.

\subsection{Radial structure of the accretion disk}

A standard way of writing the vertical hydrostatic equilibrium of a disk is 
(e.g., \citealt{1999ApJ...521..823P})
\[
\frac{\partial P}{\partial z}=-\rho \Omega_K^2 z\;.
\]
Integrated over the thin disk, the hydrostatic equilibrium defines the scale height of the disk $h$,
\emph{via} the aspect ratio $\epsilon=h/r$:
\[
\epsilon^2=\frac{P_0}{\rho_0 \Omega_K^2 r^2}\;,
\]
where quantities with the subscript 0, refer to mid-plane values. Using the perfect gas equation of 
state, the previous equation reads
\beq
\epsilon^2=\frac{k_B T_0}{\bar\mu m_p \Omega_K^2 r^2}\;,
\label{eq:epsilon}
\eeq
where $k_B$ is the Boltzmann constant, $\bar\mu$ the mean molecular weight of the gas and $m_p$ the 
proton mass.

Defining $u_0$ to be the inward radial velocity of the flow, one can simply calculate
the density as
\beq
\rho_0=\frac{\dot M_a}{4\pi r h u_0}=\frac{\dot M_a}{4\pi\Omega_K r^3}\frac{1}{m_s \epsilon^2}\;,
\label{eq:rho}
\eeq
where $m_s$ is the sonic Mach number, defined by \eq{eq:ms} in the MAES model.

Equaling the heating and cooling terms
\beq
Q^+=Q^-\;,
\label{eq:q+q-}
\eeq
and using \eq{eq:epsilon} and \eq{eq:rho}
it is possible to express all the thermodynamical quantities (in particular the midplane
temperature $T_0$ and the disk surface density $\Sigma=2\rho h$) as a function of the radius
and of the parameters of the problem, namely under the generic form
\beq
X(r)\propto {\cal A}_x(M_\star,\dot M_a, m_s,\bar\kappa, a, b)\;r^{\delta_x}\;,
\eeq
where ${\cal A}_x$ and $\delta_x$ relates to the quantity $X$ 
and depends on the problem parameters.
However, the explicit forms of these quantities can be very lengthy and 
are postponed to Appendix~\ref{app:formules}
for the sake of legibility. Note that these expressions could be directly used by
those wishing to quickly derive the structure of a JED for a given set of stellar and opacity 
parameters.

Concerning the magnetic field and as stressed in \S\ref{subsec:MAESprop}, 
the necessary condition for launching a self-collimated 
jet from a Keplerian accretion disk is the presence of a large scale vertical magnetic field 
close to equipartition ($\mu\sim 1$, \citealp{1995A&A...295..807F}). 
This condition allows us to estimate the strength of the magnetic field
\beq
B_z \simeq  0.2\ \left ( \frac{M_\star}{M_\odot}\right )^{1/4}  \left ( \frac{\dot M_a}{10^{-7} M_\odot/yr}
 \right )^{1/2} \left ( \frac{r}{1\mbox{ AU}}\right )^{-5/4+\xi/2}  \mbox{ G}\;. 
 \label{eq:Bd}
\eeq
We remind that for the specific case of our calculation, the ejection efficiency $\xi=0$.
This equation along with the set provided in Appendix~\ref{app:formules} ($T_0$, $\epsilon$ and $\Sigma$) completely determine the disk structure.

\section{Results\label{sec:results}}

The main features of jet emitting disk can now be derived from the expression 
collected in Appendix~\ref{app:formules}. In Sec.~\ref{subsec:jed_main}, 
the radial variations of the key quantities of the JED are presented. Comparison to the
standard case is also drawn.
Sec.~\ref{subsec:SED} illustrates the effect of a jet emitting region 
on the disk on the spectral energy distribution.

\begin{figure*}
\begin{center}
\includegraphics[clip=, width=16cm]{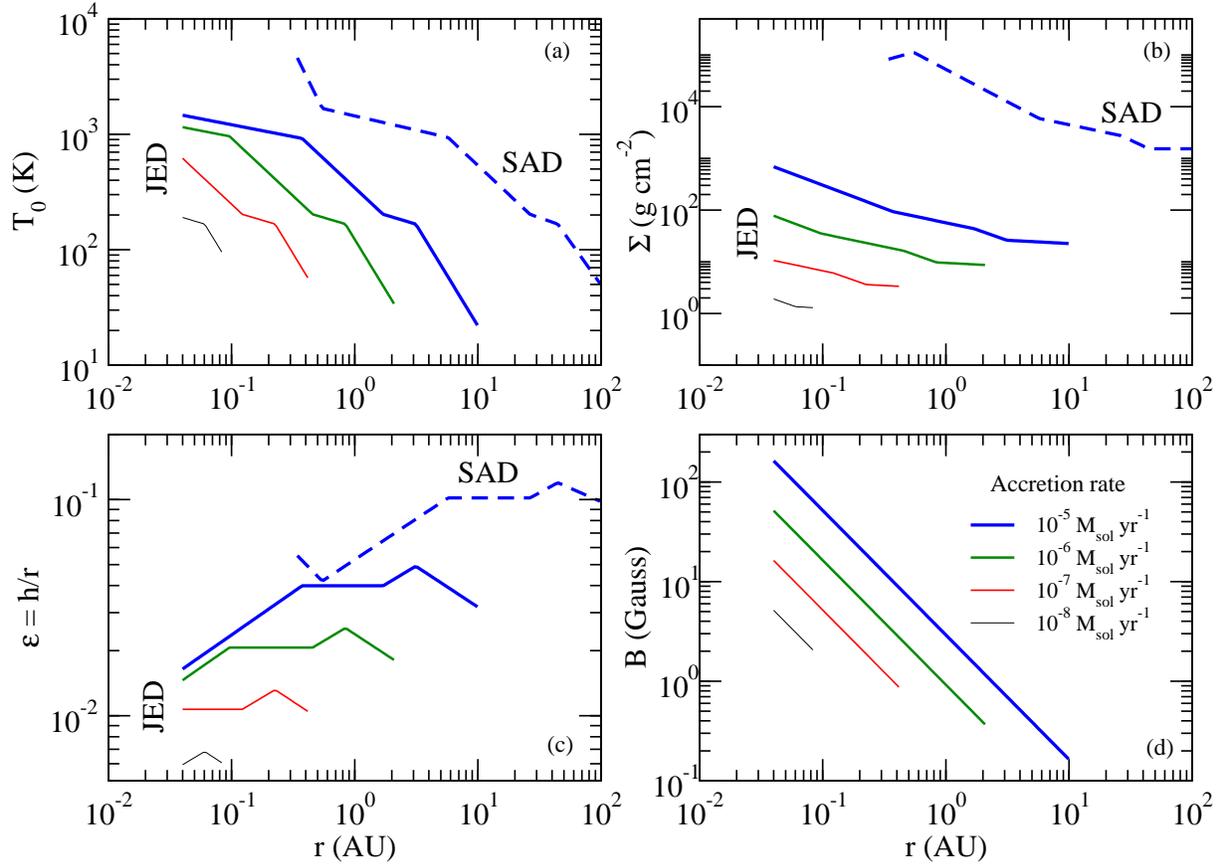}
\caption{Radial variation of key disk quantities. 
\emph{Upper-left:} mid-plane temperature. \emph{Upper-right:} surface density. \emph{Lower-left:} aspect ratio of the disk. \emph{Lower-right:} large scale vertical magnetic field intensity. Line thickness represents the increasing accretion rates considered for the JED: 
$\dot M_a=10^{-8}\;,10^{-7}\;,10^{-6}\;{\rm and}\; 10^{-5}\;\msol\;{\rm yr}^{-1}$. However, the 
SAD case is computed for $\dot M_a=10^{-5}\;\msol\;{\rm yr}^{-1}$ and $\alpha_v=0.01$ only in order to prevent overlapping curves.
\label{fig:tableau_jed}}
\end{center}
\end{figure*}

\subsection{JED: main characteristics\label{subsec:jed_main}}

In Fig.~\ref{fig:tableau_jed} are plotted in solid lines the radial dependences of  the 
central temperature (upper-left), aspect ratio (lower-left), surface density (upper-right) 
and magnetic field intensity (lower-right) of the JED, as given by Eq.~(\ref{eq:Tzero}), 
(\ref{eq:aspect-ratio}), (\ref{eq:surf-dens}) and (\ref{eq:Bd}).
The quantities---but for the magnetic field---appear  as broken power-laws, each segment corresponding to 
a given opacity regime $(\bar\kappa,a,b)$. The transition radius between two regimes of the 
Bell \& Lin opacity, say $i$ and $i+1$, are obtained by $\kappa_i(r)=\kappa_{i+1}(r)$ and solving for $r$. 

\begin{figure}
\begin{center}
\includegraphics[clip=, width=9cm]{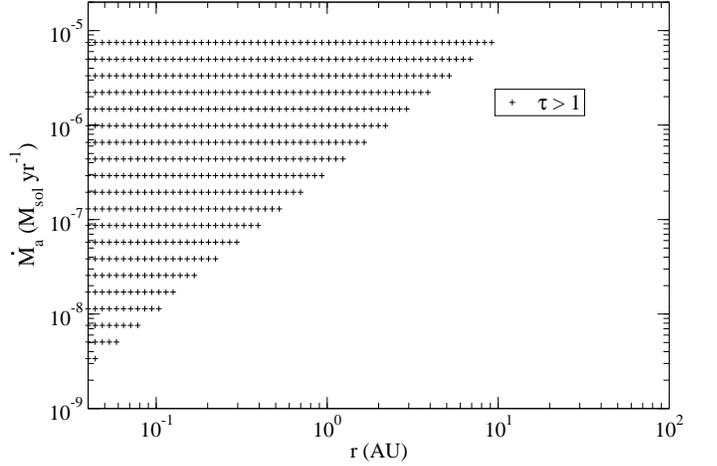}
\caption{Region of optically thick JED in the $(r,\dot M)$ plane. 
Crosses represent the validity range
of our calculation.
\label{fig:r_mdot}}
\end{center}
\end{figure}

The disk quantities have been plotted for different mass accretion rates, from
$10^{-8}$ to $10^{-5}\;\msol$~yr$^{-1}$. In agreement with intuition, the higher the accretion 
rate, the higher the temperature, surface density and disk aspect ratio. Note also that more opacity 
regimes need to be taken into account for the highest accretion rates as a result of the
increasing the temperature.
We have truncated each curve at the outer radius for which the gas becomes locally optically 
thin to its own radiation and where our description of the radiation in
\eq{eq:q-} stops being valid. But since accretion disks of CTTS are optically thick, this radius 
marks an upper limit for $r_J$. Figure~\ref{fig:r_mdot} is another representation of the latter 
point and gives in the $(r,\dot M_a)$ plane the region where the optical depth $\tau$ is greater 
than unity in a JED. There are not many sources where both the accretion rate and $r_J$ have been
estimated. This is however the case for DG Tau ($\dot M_a \sim 2\times10^{-6}\;\msol$~yr$^{-1}$, 
\citealp{2002ApJ...576..222B}) for
which \citet{2006A&A...453..785F} estimated $r_J\sim 1.5-4.5$~AU: these values fall into 
the validity range of our calculation given in Fig.~\ref{fig:r_mdot}.

Coming back to Fig.~\ref{fig:tableau_jed}, all quantities (but for the magnetic field) 
are compared to the SAD case (dashed curve). For the latter, 
we have plotted only the highest accretion rate (and use $\alpha_v=10^{-2}$) to prevent 
overlapping curves. Also not shown here, 
our SAD models are in agreement with the disk structure derived in \citet{1999ApJ...521..823P}. 
For the SAD, curves for smaller values of the accretion rate and other $\alpha_v$ 
are available in that paper.

The radial structure of a JED is found very different than that of its equivalent SAD:
at a given accretion rate, the JED is cooler (Fig.~\ref{fig:tableau_jed}a), lighter 
(Fig.~\ref{fig:tableau_jed}b) and thinner (Fig.~\ref{fig:tableau_jed}c) than the SAD. 
In particular, at a given radius, the surface density can vary by $\sim$ two orders of magnitude. 
This implies a large density jump at the transition radius $r_J$, that remains whatever the 
chosen value of the turbulent parameter $\alpha_v$.  Implications of this point are further 
discussed in \S\ref{subsec:planet}.

Turning to the magnetic field, one might object that fields of the strength given by \eq{eq:Bd} and 
illustrated in Fig.~\ref{fig:tableau_jed}d 
are impossible in accreting systems. However, the value of this magnetic field is actually 
far smaller than 
the one estimated from the interstellar magnetic field assuming either ideal MHD $B \propto n$ 
or $B \propto n^{1/2}$ \citep{1993prpl.conf..279H,1994ApJ...432..720B}. Indeed, if one takes 
the fiducial values $n \sim 1\mbox{ cm}^{-3}$ and $B\sim 4\ \mu$G observed within dense clouds 
and use the law $B\propto n^{1/2}$ \citep{1999ApJ...520..706C}, we get a magnetic field at 1~AU ranging 
from 10 to $10^3$~G (depending on the density)! Thus, the main problem is to get rid off the 
magnetic field during the infalling stage. This issue is still under debate. It seems 
nevertheless straightforward to build up accretion disks threaded by a large scale magnetic 
field of a  large amplitude from 3D collapse calculations (see e.g. \citealp{2006ApJ...641..949B}).

The observation of magnetic fields is a very difficult task and no measurements have been obtained 
so far in a disk known to drive a jet. However, \citet{2005Natur.438..466D} have managed to 
measure the magnetic field strength in the disk of FUOr, a very strong accreting object 
with $\dot M_a\sim10^{-5}\msol$~yr$^{-1}$. Using the spectro-polarimeter ESPadOnS, they found 
a value of $B\sim 1$~kG at 0.05~AU from the star. At this distance and for such an accretion 
rate, Eq.~(\ref{eq:Bd}) gives $B\sim 100$~G. The magnetic field present at the inner disk of  
FUOr is actually higher than equipartition: according to the MAES theory, no steady state 
self-collimated jet can be launched from this disk, which is indeed consistent with 
observations. If one takes this observation at face value, then one possible explanation 
is that the disk magnetic flux has been advected and compressed towards the star by the 
sudden rise of disk accretion rate. This conjecture should deserve further investigation. 
Note also that the presence of a strong (i.e. larger than equipartition)
vertical magnetic field in an accretion disc triggers non-axisymmetric
instabilities and spiral waves \citep{1992ApJ...393..708T}. If these waves
bounce back at the inner disk boundary then a standing spiral pattern is
formed and leads to the formation of a magnetized vortex (a MHD Rossby wave)
localized at the radius where the keplerian rotation coincides with the wave
frequency
\citep{1999A&A...349.1003T,2001A&A...367.1095C,2001MNRAS.323..587S}. Although
the field structure derived by \citet{2005Natur.438..466D} shows a remarkably
high degree of axisymmetry, one might indeed expect non axisymmetric
perturbations in strongly magnetized disks. However, for the physical
conditions envisioned in JEDs, namely a field close to but smaller than
equipartition, such instabilities are rather weak (see however \citealt{2002ApJ...569L.121K} for MHD instabilities at $\mu \sim 1$).

\subsection{Spectral Energy Distributions \label{subsec:SED}}

The physical properties of a jet emitting disk are rather different from that of a standard
disk. In consequence, their radiative properties should also differ and are investigated hereafter. 
Spectral energy distributions (SED) are one of the main disk diagnosis
for comparison with observations. If the radial structure of the disk does not depend too 
drastically on the irradiation of
the central object (see \S\ref{subsec:irrad}), its radiative properties certainly do 
(see \citealt{2007prpl.conf..555D} for a review). In particular, 
the simplest approach considers that the stellar illumination can create a super-heated layer at the 
surface of the disk which, in turn, 
affects the disk SED \citep{1997ApJ...490..368C,2001ApJ...560..957D}. The success of this model comes 
from both its simplicity (compared to that of a full radiative transfer treatment) and 
ability to reproduce observations of TTauris and Herbig Ae/Be stars.

However our purpose here is to compute the SED to illustrate a possible effect 
of the JED rather than reproduce specific observations.
For that reason, the irradiation of the central star on the disk will be included the very crude
fashion detailed below.
\onvire{This cannot be physically justified since the inclusion of a super-heated layer is now a well-established 
ingredient of SED modeling}. 
\onvire{We stress that using this simple configuration serves our 
purpose better in order to identify the effects of the presence of a jet emitting region in the disk.}

We will consider two components to the irradiating flux: the stellar radiation and the UV accretion
luminosity. Indeed, in the now widely accepted magnetospheric accretion model, matter is channelled along the 
stellar magnetic dipole from the disk onto the star (see \citealt{2007prpl.conf..479B} and references therein) 
where it creates the observed hot spots/rings. 
The UV radiation from these accretion shocks can then illuminate and heat the disk.
In the following, $L_{\rm UV}$ will refer to the UV luminosity of one on the two hot rings, and $L_\star$
to the stellar component. 
\citet{1998ApJ...492..323G} estimated from a sample of TTauri stars that
\[
2L_{\rm UV}\sim \frac{L_{\rm acc}}{3.5}\;,
\]
where the total accretion luminosity scales with the accretion mass rate as
\beq
L_{\rm acc}\approx \frac{GM_\star \dot M_a}{R_\star}\left(1-\frac{R_\star}{r_{\rm in}}\right)\;.
\eeq
The stellar luminosity is simply given by
\beq
L_\star=4\pi R_\star^2\sigma T_\star^4\;,
\eeq
where $R_\star$ and $T_\star$ are respectively the star radius and temperature.
The irradiation flux $Q_{\rm irr}$ is then obtained by
\beq
Q_{\rm irr}= \frac{L_\star+L_{\rm UV}}{4\pi r^2}\cos\psi\;,
\label{eq:qirr}
\eeq
with $\psi$ the angle between the incident radiation and the normal to the disk surface.
For the sake of simplicity, we will assume a fixed incident angle, $\cos\psi=0.05$ 
(i.e., incident angle $\sim$ 3~deg.): this is of course not consistent with our disk
model, as $h/r\neq\rm cst$ (Fig.~\ref{fig:tableau_jed}c), but has the benefit of giving
a quick estimate of the irradiation component.
 
With these ingredients, the disk effective temperature required by SED calculation is 
obtained by
\beq
\sigma T_{\rm eff}^4=Q^+ + Q_{\rm irr}\;.
\label{eq:Teffsed}
\eeq
The viscous heating flux $Q^+$ is given by \eq{eq:q+}---we remind here that this flux is much
smaller in a JED than in a SAD.

\begin{figure*}
\begin{center}
\includegraphics[clip=, width=15.5cm]{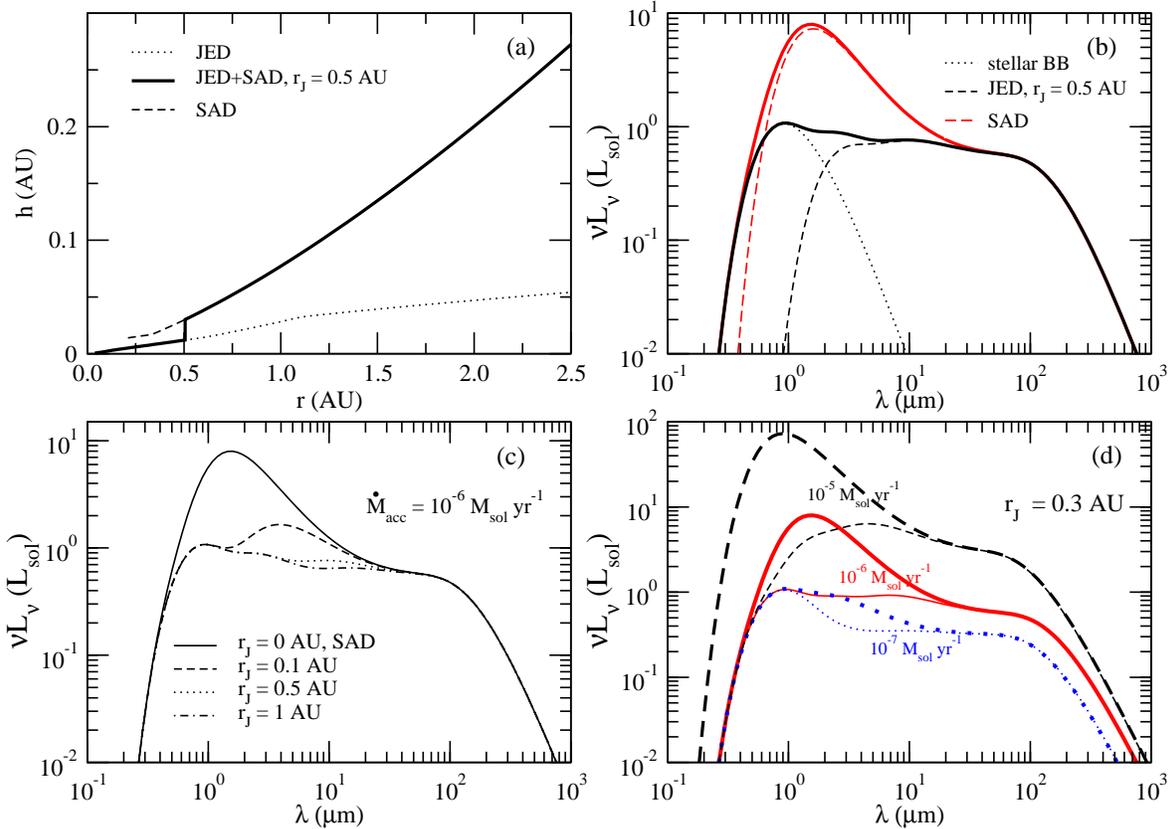}
\caption{\emph{Upper-left:} disk scale height as a function of the distance to the central object. 
The JED/SAD transition radius is $r_J=0.5$~AU. \emph{Upper-right:} pole-on viewed spectral energy of
the JED+SAD disk of (a). Dashed lines correspond to the SAD (thin red) and JED+SAD (thick black) 
configurations.
A 4000~K stellar blackbody is included (dotted line). The sum of the stellar and disk
SED is plotted in thick solid curves (red and black). \emph{Lower-left:} variation of the total JED+SAD
energy distribution with increasing transition radius. \emph{Lower-right:} influence of the accretion rate on the SED for a fixed transition radius. Thick lines correspond to the SAD alone and thin ones to the JED+SAD case.
\label{fig:tableau_sed}}
\end{center}
\end{figure*}

To actually compute the SED, we consider the disk configuration presented in Fig.~\ref{fig:schema}, where a
JED is present from $r_{\rm in}$ to $r_J$ and a SAD from $r_J$ to 
$r_{\rm out}$. We choose arbitrarily $r_J=0.5$~AU (see e.g. \citealp{2006A&A...453..785F}).
For illustrative purpose, Fig.~\ref{fig:tableau_sed}a gives the disk height scale as a function of the radius. 
The thick solid line corresponds to the disk studied here, with a transition in $r_J$. The SED obtained 
from this structure is then compared to the case where a standard disk is present on the entire
spatial domain $[r_{\rm in},r_{\rm out}]$.
Using the temperature derived in \eq{eq:Teffsed}, 
the disk SED is computed for a one solar mass star and using the simplest geometry, 
where the disk is viewed pole-on,
\begin{eqnarray}
\nonumber
\nu L_\nu^{\rm disk}&\equiv&4\pi d^2 \nu F_\nu\\\nonumber
&=&8\pi^2\nu\int_{r_{\rm in}}^{r_{\rm out}}
r B_\nu(T_{\rm eff}) \;dr \\
&=& 8\pi^2\nu \left(\int_{r_{\rm in}}^{r_J}
r B_\nu(T_{\rm JED}) \;dr + \int_{r_J}^{r_{\rm out}}
r B_\nu(T_{\rm SAD}) \;dr\right)\;.
\end{eqnarray}
In this expression, Planck's law $B_\nu$ depends of the radius through the
disk effective temperature $T$. The contribution of the central object 
to the SED is also considered, using $T_\star=4000$~K and $R_\star=2.5\;\rsol$, and assuming it 
radiates like a spherical blackbody:
\beq
\nu L_\nu^{\rm star}=4\pi^2R_\star^2 \nu B_\nu(T_\star)\;.
\eeq 

The spectral energy distribution is plotted in Fig.~\ref{fig:tableau_sed}b, from 
infrared to millimeter wavelengths. The dotted and dashed curves correspond respectively to
the stellar and disk contributions. The total SED is plotted in thick solid curves, for each
disk configuration---SAD alone or JED+SAD. The effect of a jet emitting region in the disk is 
characterized by a smaller flux and a redshifted disk SED.  
Note that \citet{2007MPLA...22.1685K} very recently reached a similar conclusion more specific 
to the case of AGN. However, although they used similar ingredients as in the present study, their outflow
model was purely phenomenological and did not rely on a physically consistent 
underlying model (such as the MAES model) for the quantification of the jet torque.
This effect increases with the size of the transition radius $r_J$: 
this is shown in  Fig.~\ref{fig:tableau_sed}c, where 
the discrepancy between the standard SED and the JED+SAD becomes larger for larger $r_J$. 
This is in agreement with intuition as it was shown in the previous section that JED were lighter 
and cooler than their equivalent SAD.

The last effect on the SED put into light in this section is shown in Fig.~\ref{fig:tableau_sed}d. For a fixed 
transition  radius $r_J=0.3$~AU, the SAD and JED+SAD spectral energy distributions are computed 
for several mass accretion rates, namely $10^{-5}$, $10^{-6}$ and $10^{-7}\msol$~yr$^{-1}$. 
For a given line pattern, the thickest line corresponds to the 
standard disk alone. For the highest accretion rate, the standard disk SED completely 
overwhelms the stellar component which explains why the thick dashed curve does not
join the others at small wavelength. 

Two interesting effects are put into light in this figure. First of all it is found
that the discrepancy between the standard and JED+SAD energy distributions is higher 
for higher accretion rates: at $10^{-6}\msol$~yr$^{-1}$ (and for that particular $r_J$), 
the standard SED is at most ten times larger than the JED+SAD one, whereas this value
falls to 2.5 at $10^{-7}\msol$~yr$^{-1}$. The second effect, linked to the first, concerns
the location of this maximum discrepency. It is found around $1.5\mu$m at 
$10^{-6}\msol$~yr$^{-1}$ and $\sim 3\mu$m at $10^{-7}\msol$~yr$^{-1}$.
\onvire{First of all, it is found that at any given wavelength (and $r_J=0.3$ AU), 
the standard flux can be as large as five times the JED+SAD one: this value varies only very 
slightly from one accretion rate to the other.  Also, it can be seen from that figure that the 
wavelength of the maximum discrepancy between the two SED increases with decreasing accretion 
rates: it is found around $\sim 2\mu$m at $10^{-5}\msol$~yr$^{-1}$, $\sim 4\mu$m at 
$10^{-6}\msol$~yr$^{-1}$ and $\sim 9\mu$m at $10^{-7}\msol$~yr$^{-1}$.} 
These two points may 
be relevant regarding observational efforts on the matter. However, let us stress one
more time that these SED are presented 
to enlight a potential effect only. Further work is required to reach a real quantification 
of this effect when taking properly into account the central object heating of the disk surface.

\section{Discussion\label{sec:discussion}}
\subsection{Illumination from the central star\label{subsec:irrad}}
In the previous section, we included irradiation of the central 
object in the calculation of the SED as it is known to have a strong influence
on the latter. However its influence onto the disk structure is less well established.
For that reason, but also for sake of simplicity, the irradiation from the central star was not 
taken into account in our calculation of the disk structure. Nevertheless, as it decreases in $r^{-2}$ compared to 
the $r^{-3}$ of the viscous heating \eq{eq:q+}, it is expected  to become 
dominant beyond a certain distance from the central object, 
geometrical effects put aside. 
\citet{1999MNRAS.303..139D} 
emphasized that the appropriate 
criterium for the disk structure to be dominated by viscous heating reads
\beq
\frac{Q_{\rm irr}}{\tau_{\rm tot}}< Q^+\;,
\label{eq:critere}
\eeq

where $Q_{\rm irr}$ is the irradiation flux coming from the star 
(see also \citealp{2006ApJ...646..275R}). 
The total optical depth of the disk $\tau_{\rm tot}$
appears as the irradiation mainly affects the surface of the disk whereas the 
viscous heating is assumed to be present on the entire disk thickness.

The irradiation flux has been determined in the previous section, by \eq{eq:qirr} and can 
readily be used with \eq{eq:q+} to check the validity of our disk structure 
calculation given by the above criterium. However, for this calculation,
we will make no assumption on the incident angle for the radiation but compute
it consistently given the disk structure we previously derived. The source of
the radiation is considered point-like. This is a reasonable assumption for the 
UV hot spot located at the end of the accretion column. This is however questionable
for the stellar component: the star is an extended source, at least for the most inner
radii. Again, for simplicity, we will consider the two components (stellar and UV)
as coming from one unique location. By changing the altitude of the spot, we should
be quite conservative in our conclusions.

\onvire{In the now widely accepted magnetospheric accretion model, matter is channelled along the stellar 
magnetic dipole from the disk onto the star (see \citealt{2007prpl.conf..479B} and references therein)
where it creates the observed hot spots/rings. The UV radiation from these accretion shocks can then 
illuminate and heat the disk. 
We will consider here that this UV luminosity is the sole source of heating of the disk. In the following,
$L_{\rm UV}$ will refer to the UV luminosity of one on the two hot rings. 
\citet{1998ApJ...492..323G} estimated from a sample of TTauri stars that
\[
2L_{\rm UV}\sim \frac{L_{\rm acc}}{3.5}\;,
\]
where the total accretion luminosity scales with the accretion mass rate as
\beq
L_{\rm acc}\approx \frac{GM_\star \dot M_a}{R_\star}\left(1-\frac{R_\star}{R_{\rm in}}\right)\;.
\eeq
}

From geometrical considerations, one gets 
\beq
\cos\psi=\frac  {h}{\sqrt{r^2+(z_{\rm sh}-h})^2}\left[\frac{d\ln h}{d\ln r}-1+\frac{z_{\rm sh}}{r}\right]\;.
\label{eq:cos}
\eeq
In the previous expression, all quantities have their usual meaning and
$z_{\rm sh}$ is the light 
spot altitude. If $z_{\rm sh}=0$, the \emph{standard} expression of \citet{2002apa..book.....F} 
is recovered.

\onvire{Using \eq{eq:qirr} and \eq{eq:cos}, we compared the irradiated flux to the local viscous heating given by 
\eq{eq:q+}.}
 Within the JED, we found that when $z_{\rm sh}$ equals zero the illumination could safely be 
discarded for any accretion rate $\dot M_a\gtrsim 3\times 10^{-7}\msol$~yr$^{-1}$
(at any radius). \onvire{It is always larger for higher accretion rates.} The situation changes when the hot spot is 
located at a higher altitude as it can strike the disk more directly in the inner regions. 
For $z_{\rm sh}=0.7R_\star$, i.e. an accretion column starting at the star at a colatitude of 45$^\circ$, 
we found that \eq{eq:critere} was not satisfied 
for $\dot M_a\lesssim 7\times10^{-7}\msol$~yr$^{-1}$.
To obtain these thresholds, we required \eq{eq:critere} to be true on the entire radius range.
Note that these accretion rates could be lowered by some extent if one only requires the condition to be
realised on \emph{most} of the radius range: it could be argued that a narrow irradiation dominated region
will not change much in the overall structure of the disk.
Basically, illumination of the JED can be neglected at any radius for $\dot M\gtrsim$ 
a few $10^{-7}\msol$~yr$^{-1}$. 
This threshold is consistent with accretion rates of sources driving powerful (detected) jets. 
Nevertheless, it would be of interest to extend this work to the lower accretion rates
by including irradiation in the heating term\footnote{This can only be done iteratively 
(hence, not analytically) as the knowledge
of the disk structure is required for the determination of $\cos\psi$.}.
From these considerations, we are left confident with the validity of 
our very simplified analytical description, at least for $\dot M\gtrsim$ a few $10^{-7}\msol$~yr$^{-1}$.

\subsection{Implications for planet formation and migration\label{subsec:planet}}
The initial conditions for planet formation depend on the disk physical 
properties. It has been shown in the previous section 
that the latter are very different from a JED to a 
SAD. We discuss, hereafter, two of the possible consequences that an accretion structure
such as that of Fig.~\ref{fig:schema} may have on planet formation. 

\citet{1996ApJ...457..355G} was the first to present the idea 
of \emph{layered accretion disks}
where the upper part of the disk is ionized---via collisions, cosmic rays or X-rays---and some embedded 
inner part stays neutral. The latter, termed \emph{dead zone} is then
decoupled to the magnetic field, hence to the MRI induced turbulence, and remains quiescent. 
Angular momentum is very poorly transported outward in the dead zone and no accretion occurs in 
this region of the disk.

We have shown in the previous section that jet emitting disks were both thinner and 
lighter than standard disks. As a consequence, their are likely to be more ionized than
SAD as X-ray radiation from the central star and cosmic rays 
should deeperly penetrate them. The calculation of the actual ionization degree of JED 
is postponed to a forthcoming study, which should give a definite answer to the possibility
of a dead zone in a JED. However, it is of interest to briefly mention a few issues related
to that matter. 

Both low and high mass planet formation start with the growth of planetesimals and the 
necessity of dust settling and agglomeration. The effect of turbulence on agglomeration is still 
being debated. Indeed, it might help to form planetesimals by trapping dust within turbulent vortices 
(e.g. \citealp{1995A&A...295L...1B}). On the other hand, recent numerical simulations of a fully turbulent  disk showed that only the larger grains (with typical sizes from 1 to 10~cm) do settle towards the midplane whereas the smaller ones stay in suspension in the disk \citep{2006A&A...452..751F}. The thickness of the dust subdisk was also found to be rather large ($\sim 0.2h$). These authors also showed that the dust subdisk is thinner (more compact) when a dead zone is taken into account in the simulations. 
Along the same line, \citet{2007ApJ...654L.159C} stressed that the efficiency of dust settling and
coagulation in the presence of a dead zone was enhanced compared to that of a ionized (i.e., turbulent)
disk. From those considerations, layered accretion disks appear to be a privileged location for grain growth and planetesimal formation. If JEDs are indeed fully ionized hence turbulent, then it is unlikely that they can host the earliest stages of planet formation.

On the other hand, the maximum value of the JED-SAD transition radius $r_J$, as determined by the transition to the optically thin JED regime, is rather small (see Fig.~\ref{fig:r_mdot}). It reaches $\sim 10$ AU only for high accretion rates, found only in embedded sources (Class 0 or I). For accretion rates of some $10^{-8}$ M$_\odot$~yr$^{-1}$, more typical of CTTS, $r_J$ cannot exceed 0.1-0.3 AU. As the JED has a rather low column density $\Sigma$ (see Fig.~\ref{fig:tableau_jed}b), 
it provides very unideal conditions for the core accretion process leading to massive planet formation.

Planet migration in a gaseous disk surrounding a central star  \citep{1980ApJ...241..425G} 
has been a long standing issue regarding planetary formation theories. 
The transfer of angular momentum between the planet and the gaseous disk generally results in the inward motion 
of the planet. This general mechanism affects both low (type~I migration) and high mass planets  
(type~II migration)---see \citep{2007astro.ph..1485A} for a review---and is generally invoked to explain 
the existence of hot Jupiters. However, the existence of giant planets \emph{far} from their central star 
(a few AU) is a real issue. Indeed, type~I migration is known to be a very fast process and should not allow 
enough time for core accretion (build up of a massive solid core), i.e. for the existence of giant planets 
at such distances. Hence, finding a way to slow down or halt type~I migration is of importance if one is 
to understand planetary system formation. 
 
In that respect, \citet{2006ApJ...642..478M} studied the effect of a sudden radial  surface density decrease (going inwards) in the disk on type~I planet migration. These authors found that  such a density jump could indeed trap a protoplanet at the location of the transition by a balance between
 the corotation and Lindbald torques. As shown previously, the surface density of a SAD is always larger 
than that of a JED for a given accretion rate (see Fig.~\ref{fig:tableau_jed}b). Hence,
a transition from an outer SAD to an inner JED, as described in Sec.~\ref{subsec:transition}, would
naturally provide the surface density jump necessary to a \emph{planet trap}. It is therefore likely that 
planetesimals should first form in the outer, denser SAD and then migrate towards the center to be 
halted at the JED-SAD transition. 

One must however be cautious as in \citet{2006ApJ...642..478M} the physics of the JED and, in particular the presence of the large scale magnetic field, was not taken into account. It would be useful, as a second step, to include this critical ingredient and determine precisely how will the JED be itself affected by the presence of a protoplanet.

\subsection{The $\dot M_a$--$M_\star$ relationship in the JED model}

One of the striking observational properties of T~Tauri stars and brown dwarfs is the 
apparent steep correlation between the mass of the central object and the disk accretion rate,
namely 
\beq
\dot M_a \propto M_\star^\alpha
\eeq
with $\alpha$ lying between 1 and 2 \citep{2003ApJ...592..266M,2004AJ....128.1294C, 2006ApJ...639L..83A,
2006ApJ...648..484H}. 
Though seemingly very robust observationally -- four orders of magnitude in accretion rates and 
two in mass--, the physical
origin of this relation is still debated. \citet{2005ApJ...622L..61P} evoked the possibility of a star-disk
accreting from a large scale envelope at the Bondi-Hoyle rate, giving precisely $\dot M_a \propto M_\star^2$.
However, this scenario does not take into account the angular momentum of the infalling gas. In an other approach,
\citet{2006ApJ...645L..69D} showed how taking into account the imprint of the physical properties of the 
parent core onto the star-disk system lead to $\dot M_a \propto M_\star^{1.8}$. \citet{2006ApJ...648..484H} explored
several mechanisms but concluded on a word of caution: these authors noted that, if the bulk of the data is
well represented by $\dot M_a \propto M_\star^2$, the TTauri with the highest accretion rates appear to have
 $\dot M_a \propto M_\star$. In that sense, \citet{2006ApJ...648..484H} stated that it may not be
relevant to look for a universal $\dot M_a$--$M_\star$ relationship, as different mass regimes may have
different dominating processes. Nevertheless, we believe it is of interest to see what  
$\dot M_a$--$M_\star$ relationship is predicted by the JED model. This is done below, using very simple
arguments.

The disk accretion rate $\dot M_a$, as derived from observations, is actually indicative of the mass flow at the inner disk regions, hence representative of the JED rather than of the outer SAD. But in a JED accretion is achieved through angular momentum removal in jets. This is made possible only because of the presence of a large scale disk magnetic field $B_{\rm disk}$. As argued in Sect.~2.2 this field is probably the parent cloud core magnetic field $B_{\rm core}$ that has been advected and concentrated by the infalling material during the collapse. Such a collapse is induced when the ratio of the magnetic flux to the total mass $\Phi/M_{\rm core}$ reaches a critical value. This translates into
\beq
B_{\rm core} \propto  M_{\rm core} 
\eeq
as an initial condition. Now, assuming that the ratio $\Phi/M$ remains constant (or varies only slowly) during the collapse, one gets  
\beq
B_{\rm disk} \propto  M_\star
\label{eq:Bdisk}
\eeq
since most of the infalling cloud core mass ends up in the central star (ejection will not change the 
relation $ M_\star \propto M_{\rm core}$). Note that another way to obtain a relation of this kind 
is by writing $B_{\rm disk} \propto B_{\rm core}^\eta$, where the value of the exponent $\eta$ 
should be provided by full 3D collapse calculations. However, we expect it to be close to unity 
which would lead to Eq.~(\ref{eq:Bdisk}) as well.

The disk magnetic field required in a JED is given by Eq.~(\ref{eq:Bd}) and leads to
\beq
B_{\rm disk}\propto \dot M_a^{1/2} M_\star^{1/4}\;.
\eeq
Inserting this relation into Eq.~(\ref{eq:Bdisk}) provides then the $\dot M_a$--$M_\star$
relationship expected in JEDs, namely
\beq
\dot M_a\propto M_\star^{3/2}\;.
\eeq

If our crude derivation gives a result consistent with the observations, one must however
remain cautious: for a given stellar mass $M_\star$ there is observationnaly a large dispersion 
in  $\dot M_a$. 
This has lead \citet{2006MNRAS.370L..10C} to point out a possible 
incompleteness of the samples used, that may have lead to such an apparent correlation. 

\section{Conclusions}
There has been growing observational evidence that the bipolar ejections of 
matter occurring during the formation of a star are directly linked to the accretion 
process. From that point of view, jets should have an influence on the
structure of the region of the disk that is launching them. 

Using the framework of the Magnetised Accretion Ejection Structures, we derived
the radial structure of such a Jet Emitting Disk and compared it to the 
equivalent Standard Accretion Disk. We found that a JED is cooler, thinner,
and lighter than a SAD at the same accretion rate. Invoking the idea of a 
radial transition from an inner JED to an outer SAD, we have also shown 
in a very crude approach that such a transition has an effect on the
Spectral Energy Distribution: the flux coming from a JED is smaller and
redshifted compared to the SAD. A more detailed study is however needed to 
provide realistic observational predictions.

Several implications on planet formation have also been drawn from the existence of JEDs.
On the one hand, and although this needs confirmation,  dead zones may not exist in JEDs as their
small thickness and density will favor ionization. Recent numerical works show that dust settling
could be made more difficult in the absence of a dead zone. This questions the 
possibility for a JED to host the earliest stages of planet formation. On the other hand, the surface density jump occurring in a JED/SAD transition could serve as a planet 
trap and halt type I migration.

Protostellar jets have been observed and theoretically studied by 
the magneto-centrifugal approach (among others) for almost thirty years.
Nevertheless, little attention has been paid to their potential
effects on the disk structure. In this paper, we stress that these effects
indeed exist and further work is to be undertaken for better quantification
and observational perspective.    

\begin{acknowledgements}
The authors wish to thank the referee, C. Dullemond, for his many
suggestions that helped us improving the manuscript. Also, C.C. wishes to thanks 
Dr. Fabio De Colle for useful discussions on irradiated disks. 
The present work was supported in part by the European Community Marie Curie 
Actions - Human Resource and Mobility within the JETSET (Jet Simulations, Experiments 
and Theory) network under contract MRTN-CT-2004 005592.

\end{acknowledgements}

\begin{appendix}

\section{MAES global energy budget\label{app:bilan}}
The global energy budget of a magnetized accretion-ejection structure writes 
\citep{1993A&A...276..625F,2000A&A...361.1178C}
\beq
P_{\rm acc} = 2P_{\rm jet} + 2P_{\rm rad}
\eeq
where the accretion power, namely the power released by the accreting flow in the disk
\beq
P_{\rm acc} \simeq \frac{GM_\star \dot M_{a,J}}{2r_{in}} \left [ \left (
\frac{r_{in}}{r_J} \right )^\xi - \frac{r_{in}}{r_J} \right ]
\eeq
is obtained by computing the difference between the mechanical power that comes in at 
$r_J$ and goes out at $r_{in}$. Note that $\dot M_{a,J}$ is a constant and must be understood here as the accretion rate feeding the JED at $r_J$ (within the JED, the disk accretion rates varies as $\dot M_a \propto r^\xi$).  All available power $P_{\rm acc}$ is thus shared between radiative losses at the disk surfaces $P_{\rm rad}$ and jet power $P_{\rm jet}$. This last term marks obviously the difference with the global energy budget of a SAD. In the latter, all accretion power is released as radiation 
whereas JEDs also power jets. This last  contribution to the budget is actually the easiest 
to compute as it is the flux of energy that leaves the two disk surfaces namely
\beq
P_{\rm jet} = \int \left [ {\bf S}_{MHD} + \rho {\cal E}  {\bf u}_p\right ] \cdot {\bf dS}
\eeq
where ${\cal E}= u^2/2 + \Phi_G + H$ is the non-magnetic specific energy ($\Phi_G$ is the gravitational potential and $H$ the enthalpy), ${\bf S}_{MHD} = - \Omega_F r B_\phi \bf{B}_p/\mu_o$ is the MHD Poynting vector ($\Omega_F \simeq \Omega_K$ is the angular velocity of the magnetic field 
line) and ${\bf dS}= dS {\bf n}$ with $dS$ the elementary disk surface and ${\bf n}$ the unit vector normal to the disk surface. For magnetically driven jets from keplerian accretion disks, the dominant term is the magnetic contribution due to the MHD Poynting vector. Thus, the energy flux carried away by the jets depends directly on the toroidal field at the disk surface $B_\phi^+$ and writes 
\beq
\frac{2P_{\rm jet}}{P_{\rm acc}} =  \frac{\Lambda}{1+ \Lambda} \left| \frac{B^+_{\phi}}{qB_z}\right |  
\eeq
where $\Lambda \sim 1/\epsilon \gg 1$ is the ratio of the jet to viscous torques and $q$ 
the magnetic shear parameter. Most MAES solutions were found with $| B^+_{\phi} | \simeq q B_z$, 
which translates into $P_{\rm acc} \simeq 2P_{\rm jet}$. Thus, most available power feeds 
the jets and energy conservation tells that only a fraction of it, namely 
 \beq 
\frac{2P_{\rm rad}}{P_{\rm acc}} \simeq  \frac{1}{1+ \Lambda} \simeq \varepsilon
\eeq
is dissipated within the JED. This justifies the use of Eq.~(\ref{eq:f}). Note however that some 
solutions were found with $| B^+_{\phi}|/qB_z$ as low as 0.5, which shows that JEDs may also, 
under certain circumstances, produce some luminosity. But as a first step to illustrate the 
effect of JEDs in the center of YSO accretion disks, we have disregarded these peculiar solutions 
(see also discussion in \citealp{2006A&A...447..813F}).

\section{JED key quantities\label{app:formules}}

In this appendix, we give the detailed analytical expressions of the main
disk quantities as calculated in the MAES context. We only focus on the radial 
dependence of each quantity. In these expressions, the
Rosseland mean opacity is taken under the form $\kappa=\bar\kappa\rho^a T^b$,
where the parameters $\bar\kappa$, $a$ and $b$ are to be determined depending
on the considered opacity regime. 
In the following, $\sigma$ is the Stefan-Boltzmann constant, $M_\star$ and $\dot M_a$ 
respectively represent the mass of the central object and the mass accretion rate, $\bar\mu = 2$ 
\citep{1999ApJ...521..823P} is the mean molecular weight of the gas, $m_p$ the proton mass and $k_B$ 
the Boltzmann constant.
The parameter $m_s$ is related to the MAES model (see Sec.~\ref{sec:MAES}) and is found 
to be close to unity in most solutions of the problem. Using Eqs.~(\ref{eq:epsilon}) and (\ref{eq:rho}) to rewrite 
\eq{eq:q+q-} from Eqs.~(\ref{eq:q+}) and (\ref{eq:q-}), the radial dependence of the mid-plane temperature 
is expressed as
\beq
T_0(r)={\cal A}_T\;r^{-\Gamma\left[1+5(a+1)/2\right]}\;,
\label{eq:Tzero}
\eeq
where $\Gamma\equiv 1/(4+a-b)$ and ${\cal A}_T$ is dependent of the parameters of the problem,
\[
{\cal A}_T=\left(\frac{\bar\kappa}{8\pi\sigma}\right)^\Gamma 
\left(\frac{1}{4\pi m_s}\right)^{\Gamma(a+1)}
(GM_\star)^{\Gamma(a+1)/2} \dot M_a^{\Gamma(a+2)}
\left(\frac{\bar\mu m_p}{k_B}\right)^{a\Gamma}\;.
\]
Plugging
\eq{eq:Tzero} into \eq{eq:epsilon}, the disk aspect ratio $\epsilon=h/r$ is then directly obtained under the form
\beq
\epsilon(r)={\cal A}_\epsilon\;r^{\frac{1-\Gamma}{2}-\frac{5\Gamma(a+1)}{4}}\;,
\label{eq:aspect-ratio}
\eeq
\[
{\cal A}_\epsilon=\left(\frac{\bar\kappa}{8\pi\sigma}\right)^{\Gamma/2} 
\left(\frac{1}{4\pi m_s}\right)^{\frac{\Gamma(a+1)}{2}}
(GM_\star)^{\frac{\Gamma(a+1)}{4}-\frac{1}{2}} \dot M_a^{\frac{\Gamma(a+2)}{2}}
\left(\frac{\bar\mu m_p}{k_B}\right)^{\frac{a\Gamma}{2}-\frac{1}{2}}\;.
\]
The disk mass density can be calculated in a similar fashion, 
(using Eqs.~(\ref{eq:aspect-ratio}) and (\ref{eq:rho}) together), but we  
rather give the expression of the surface density as it appears to be 
a more relevant quantity regarding accretion disk studies. The expression 
of $\Sigma(r)$ is calculated using 
$\Sigma(r)=\int_{-h}^{+h} \rho(r,z)dz \sim 2 \rho_0 h=2 \rho_0(r) \epsilon(r) r$ 
and reads
\beq
\Sigma(r)={\cal A}_\Sigma\; r^{\frac{\Gamma}{2}+\frac{5\Gamma(a+1)}{4}-1}\;,
\label{eq:surf-dens}
\eeq
with
\[
{\cal A}_\Sigma=\left(\frac{\bar\kappa}{8\pi\sigma}\right)^{-\Gamma} 
\left(\frac{1}{4\pi m_s}\right)^{1-\frac{\Gamma(a+1)}{2}}
(GM_\star)^{\frac{-\Gamma(a+1)}{4}} \dot M_a^{1-\frac{\Gamma(a+2)}{2}}
\left(\frac{\bar\mu m_p}{k_B}\right)^{\frac{1}{2}-\frac{a\Gamma}{2}}\;.
\]

\end{appendix}

\bibliography{soumis2}

\end{document}